\documentclass[aps,prb,twocolumn,floatfix,superscriptaddress,longbibliography,nofootinbib,10pt]{revtex4-2}

\usepackage{graphicx, color, soul}
\usepackage{amsmath, amsfonts, amssymb, mathrsfs, dsfont, mathtools}

\usepackage[dvipsnames, pdftex, table]{xcolor}
\usepackage[normalem]{ulem} % needed for \scrap-command
\usepackage[mathscr]{eucal}
\usepackage{bm}
\usepackage{tabularx, booktabs}
\usepackage{multirow}
\usepackage{physics}
\usepackage{bbold}
\usepackage{comment}
\usepackage{wasysym}
\usepackage{tikz}    % For drawing shapes
\usepackage{stackengine}% for stacking symbols
\usepackage{csquotes}
\usepackage[colorlinks, breaklinks, 
            linkcolor=Blue,
            citecolor=Blue,
            urlcolor=Blue]{hyperref}
            
\usepackage[all]{hypcap} 
\usepackage{enumitem}
\usepackage{placeins}
\usepackage[english]{babel}

\usepackage{xcolor}

\definecolor{mutedblue}{RGB}{45,85,135}   % balanced deep
\stackMath
\date{\today}

\begin{document}
\title{Frustrated magnetic order in hybrid Kitaev spin-orbital models}
%\title{Hybrid Kitaev spin-orbital liquid models: novel phases, magnetic fragmentation and exact solution}
%\title{Ground states of exactly solvable bilayer Kitaev honeycomb model}

\author{Ivan Dutta}
%\email{ivan.dutta@niser.ac.in}
\affiliation{National Institute of Science Education and Research, Jatni, 752050, India}
\affiliation{Homi Bhabha National Institute, Training School Complex, Anushakti Nagar, Mumbai 400094, India}
\author{Aayush Vijayvargia}
\affiliation{Department of Physics, Arizona State University, Tempe, Arizona 85287, USA}
\author{Anamitra Mukherjee}
\affiliation{National Institute of Science Education and Research, Jatni, 752050, India}
\affiliation{Homi Bhabha National Institute, Training School Complex, Anushakti Nagar, Mumbai 400094, India}
\author{Onur Erten}
\affiliation{Department of Physics, Arizona State University, Tempe, Arizona 85287, USA}
\author{Kush Saha}
%\email{kush.saha@niser.ac.in}
\affiliation{National Institute of Science Education and Research, Jatni, 752050, India}
\affiliation{Homi Bhabha National Institute, Training School Complex, Anushakti Nagar, Mumbai 400094, India}
\affiliation{Max-Planck Institute for the Physics of Complex Systems, Noethnitzer Str. 38, 01187, Dresden, Germany}

\begin{abstract}

%We study hybrid quantum spin-liquid systems that combine two independently exactly solvable models: the Kitaev honeycomb model and its spin–orbital generalization, the Yao–Lee model. While such models are exactly solvable in isolation, real materials often host multiple competing interactions that break solvability and generate new phases. Motivated by recent microscopic proposals, we investigate what happens when Kitaev- and Yao–Lee–type couplings coexist on distinct lattices with the same coordination number, focusing on honeycomb and square-lattice geometries.

%Using a self-consistent perturbative mean-field approach, we show that in the strong Kitaev regime the system exhibits magnetic fragmentation: conventional magnetic order develops in the spin sector, while the orbital sector retains its topological spin-liquid character. In the hybrid Yao–Lee–square-lattice system, we find a rich evolution of Majorana band structures, including Dirac cones and Lifshitz transitions. Notably, when the Yao–Lee and square-lattice couplings are equal in magnitude and opposite in sign, the model becomes exactly solvable again, featuring a single itinerant Majorana mode.

%These results highlight how combining distinct spin-liquid Hamiltonians can generate novel emergent phases, offering a new route for exploring complex quantum states in correlated materials.

Spin-orbital generalization of Kitaev model provides a robust extension to the original Kitaev model.
However, real materials often exhibit competing interactions that break exact solvability which can give rise to new phases.
Motivated by recent microscopic proposals of coexisting Yao-Lee and Kitaev couplings, we investigate the fate of the ground state when two independent exactly solvable spin liquid Hamiltonians each originally formulated on different lattice geometries are combined on a common lattice environment. 
We first focus on the hybrid Kitaev's honeycomb and square-lattice model. Using self-consistent mean-field analysis and perturbative calculation, we show that the strong-Kitaev regime yields magnetic order in the spin sector, while the orbital sector retains its topological order. We further analyze the hybridization of the Yao-Lee and square-lattice models and find that the model exhibits a rich evolution of Majorana Dirac bands and Lifshitz transitions. Remarkably, when the Yao-Lee and square-lattice couplings are equal and opposite, the model restores its exact solvability with a single itinerant Majorana flavor.
These results demonstrate that hybrid spin liquid platforms may host various emergent phases beyond conventional exactly solvable limits.

\end{abstract}
\maketitle

\section{Introduction}
Since its introduction, the Kitaev honeycomb model ~\cite{Kitaev_Model, HYKee2025} continues to inspire new directions in the study of quantum spin liquids (QSLs)~\cite{Zhou_RMP, Roderich_Annual_Review, Wen_2017, hermanns2018,Senthil_Science, Balents2010_Nature, Savary_2017, Moessner_Moore_2021, Mandal_Kitaev}. 
 While the ideal model is exactly solvable, real candidate materials such as $\alpha$-RuCl$_3$~\cite{Takagi2019_Rucl3}, iridates ~\cite{hwan2015_nature, Kitagawa2018_nature}, or van der Waals (vdW) materials~\cite{Chris_2020_VdW, blei2021_AIP} inevitably host additional symmetry-allowed interactions such as Heisenberg and off-diagonal exchange terms~\cite{BeyondKitaevInt1, BeyondKitaevInt2, BeyondKitaevInt3}. These additional interactions introduce quantum fluctuations that tend to destabilize the spin liquid phase and may give rise to conventional magnetic order~\cite{Rau_AnnRev2016,Vijayvargia_PRB_2024}.

This has motivated the search for more robust exactly soluble models that can accommodate such perturbations without losing the exact solvability. An important step in this direction is the Yao–Lee model~\cite{Wu_Gamma_PRB,YaoLee2011}, which emerges from Kugel–Khomskii interactions~\cite{Kugel1977, Kugel1982}. By enlarging the local Hilbert space through the inclusion of orbital degrees of freedom, this model allows for a broader class of interactions that commute with the flux operators, thereby preserving exact solvability. As a result, it provides a more flexible platform to explore the interplay between spin liquid behavior and additional interactions~\cite{Yang_2007, Yao_2007_PRL, Yao_PRL_2009, Saptarshi_2009, baskaran2009exact, Wu_Gamma_PRB, Tikhonov_PRL_2010, Chua_PRB_2011, Furusaki_PRB_2012, Chulliparambil_2020, Akram_Vison_2023, Keskiner_PRB_2023, Chulliparambil_PRB2021, Dutta2025, Carvalho2018, Poliakov2024,Corboz2012, KESKINER2025,Vijayvargia_arXiv2025,Vijayvargia_PRR_2023,Urban2020, Nica2023_NPJ}. 
Recent microscopic proposals for realizing Yao-Lee interactions show that additional Kitaev and Heisenberg terms can naturally appear in the same spin-orbital setting~\cite{Kee2025}. Related studies have shown that such perturbations can produce magnetically ordered phases while preserving aspects of the underlying liquid sector~\cite{akram2025, Fornoville2025}. %Motivated by these developments, we ask a broader question: what happens when two independent Hamiltonians, each originally formulated on different lattice geometries and supporting an exactly solvable spin-liquid ground state, are hybridized on a lattice with a shared coordination number? {\bf [OE: shared?]}
Motivated by these developments, we ask what happens when two independently solvable Kitaev-type Hamiltonians, originally formulated on different lattice geometries, are placed on a common coordination-four lattice. 
We first study a combined model consisting of the Kitaev honeycomb interaction in the orbital sector and an SO(2)-symmetric square-lattice spin-orbital generalization. 
Away from the decoupled limits, the model is no longer exactly solvable, and we analyze it using perturbation theory and self-consistent Majorana mean-field theory. 
In the strong-Kitaev limit, an infinitesimal square-lattice spin-orbital coupling lifts the degeneracy of the decoupled spin sector and induces magnetic order. 
Depending on the relative signs of the Kitaev and square-lattice couplings, the resulting ordered states are ferromagnetic, antiferromagnetic, or spiral, as summarized in Fig.~\ref{fig:lattice_geometry}c. 
The magnetic order appears in the spin sector, while the orbital sector retains the conserved Kitaev flux structure and liquid character.

We then consider the analogous combination of the SO(3)-symmetric Yao-Lee model and the SO(2)-symmetric square-lattice model. In contrast to the Kitaev-square case, the mean-field solution does not develop magnetic order. Instead, the competition between the three Majorana flavors of the Yao-Lee model and the two Majorana flavors of the square-lattice model reconstructs the low-energy band structure, producing a sequence of Dirac-node motions and Lifshitz transitions~\cite{Lifshitz1, Lifshitz2}. 
At the special point where the Yao-Lee and square-lattice couplings are equal in magnitude and opposite in sign, the model becomes exactly solvable again and supports a single itinerant Majorana flavor.

The rest of the paper is organized as follows. In Section~\ref{sec:Kitaev Square}A, we introduce the hybrid Kitaev–square model, discuss its plaquette algebra, and present the corresponding Majorana representation. The perturbative analysis and mean-field results for the two choices of Kitaev coupling are then presented in Sections~\ref{sec:Kitaev Square}B, \ref{sec:Kitaev Square}C, and \ref{sec:Kitaev Square}D, respectively. Section~\ref{sec:Kitaev_Yao_Lee} is devoted to the Yao-Lee square model, together with its mean-field phase diagram in Sections~\ref{sec:Kitaev_Yao_Lee}A and \ref{sec:Kitaev_Yao_Lee}B. Finally, we summarize our main results and outline possible directions for future work in Section~\ref{sec:Conclusions}.

% The generalized Kitaev model, specifically the Yao-Lee model defined on honeycomb lattice with coupled spin and orbital degrees of freedom has received
% tremendous attention in recent years\cite{}. While the exact solvability of spin-1/2 Kitaev model is limited to only a small set of interactions, the Yao-Lee model retains exact solvability for a broarder class  of interactions. Additionally, even with interactions that breaks the exact solvability of the Yao-Lee model, the presence of additional degrees of freedom allows a rich variety of emergent phases. For example, Yao-Lee model with added Kitaev and Heisenberg interactions allows distinct magnetic phases along with spin-liquid phases for different regime of interaction\cite{akram}.

% Partly motivated by these ideas, we extend the previous studies by incorporating a different class of interactions in both the standard Kitaev honeycomb model with additional degrees of freedom and the Yao–Lee model. In particular, we show that
%%%%%%%%%%%%%%%%%%%%%%%%%%%%%%%%%%%%%%%%%%%%%%%%%%%%%%%%%%%%
\section{\label{Section_2} Kitaev - square model}\label{sec:Kitaev Square}
%%%%%%%%%%%%%%%%%%%%%%%%%%%%%%%%%%%%%%%%%%%%%%%%%%%%%%%%%%%%
%%%%%%%%%%%%%%%%%%%%
\subsection{Model Hamiltonian}\label{subsec:modelks}
\begin{figure}
    \centering
    \includegraphics[width=0.97\linewidth]{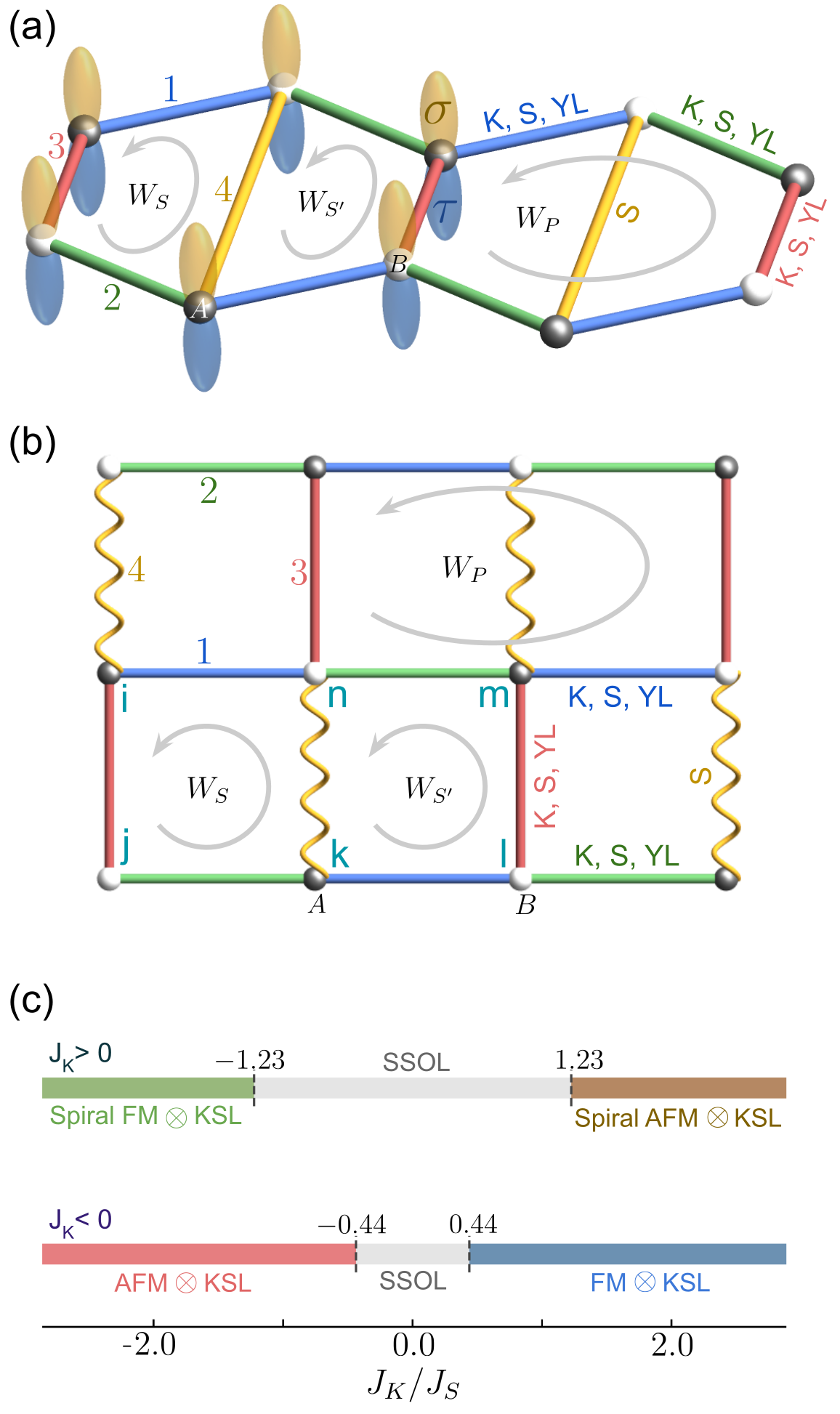}
    \caption{(a) Schematic of the modified honeycomb lattice with coordination number-$4$. Each lattice site (with sublattice A and B) hosts independent spin $\sigma$ and orbital $\tau$ degrees of freedom. The Kitaev (K), square (S), and Yao-Lee (YL) interactions are assigned to specific bond types, distinctly labeled 1, 2, 3, and 4. The operators $W_p$ and $W_{s,s'}$ denote the counterclockwise defined conserved hexagonal and square plaquettes, respectively. (b) A geometrically equivalent square-lattice representation, illustrating that all the aforementioned definition remains same. (c) The phase diagram of the Kitaev-square model for their relative coupling strength. Here SSOL and KSL denote the square spin-orbital liquid phase and Kitaev spin liquid phase respectively.}
    \label{fig:lattice_geometry}
\end{figure}

%%%%%%%%%%%%%%%%%%%%
We consider a two-dimensional lattice defined on a modified honeycomb geometry with coordination number four, as illustrated in Fig.~\ref{fig:lattice_geometry}(a). Both Kitaev ($H_K$) and square-lattice generalizations of the Kitaev model ($H_S$) Hamiltonians can be defined on this lattice and the combined Hamiltonian reads off
\begin{equation}
H_{ks}=H_{K}+H_{S}.\label{eq:HKS}
\end{equation}

The same connectivity can be represented either as a modified honeycomb lattice or as a square-lattice, as shown in Fig.~\ref{fig:lattice_geometry}(b). 
In the square-lattice representation, the square-lattice Hamiltonian acts on all four bond types, while the Kitaev interaction occupies a brick-wall subset of three bonds. We use these two representations interchangeably, since they correspond to the same bond connectivity and operator assignment.

%%%%%%%%%%%%%%%%%%%%%%%%%%%%%%%%%%%%%%%%%%%%%%%%%%%%%%%%
%%%%%%%%%%%%%%%%%%%%%%%%%%%%%%%%%%%%%%%%%%%%%%%%%%%%%%%%

At each lattice site, we consider two independent quantum degrees of freedom: a spin-1/2 and a pseudospin-1/2 (orbital) sector, described by two mutually independent sets of Pauli matrices 
$\sigma$ and $\tau$, respectively. Together, they span a four-dimensional local Hilbert space at each site. We note that, the roles of the spin and orbital degrees of freedom are interchangeable in this construction, as discussed in Ref.~\cite{Kee2025}. The  individual Kitaev and square-lattice Hamiltonians~\cite{Chulliparambil_2020} in terms of these degrees of freedom are defined as
\begin{eqnarray}
H_{K}&=&J_K\sum_{\langle ij \rangle_\alpha} (\tau_i^\alpha \tau_j^\alpha) \label{eqK}\\
% H_{YL-S}&=&J_{Y}\sum_{\langle ij \rangle_{\alpha}}(\sigma_i \cdot \sigma_j)(\tau_i^\alpha \tau_j^\alpha) \label{eqYL}\\
H_{S}&=&J_S\sum_{\langle ij \rangle_{{\alpha'}}}(\sigma_i^x \sigma_j^x+\sigma_i^y \sigma_j^y)(\tau_i^{\alpha'} \tau_j^{\alpha'}). \label{eqSqL}
\end{eqnarray}

Here $\alpha=\{1,2,3\}$ labels the three standard Honeycomb bond directions for $H_K$, while ${\alpha'}= \{1,2,3,4\}$ is used to include the fourth bond (yellow colored) for $H_S$. It is worth mentioning that unlike the standard Kitaev honeycomb model, which is typically formulated in terms of spin-1/2 degrees of freedom, $H_K$ in the present construction acts exclusively on the orbital ($\tau$) sector. The Hamiltonian $H_S$ is an SO(2)-symmetric square-lattice spin-orbital generalization of the Kitaev model, with the SO(2) symmetry corresponding to global rotations in the $\sigma^x$-$\sigma^y$ plane.

In the decoupled limits, both $H_K$ and $H_S$ are exactly solvable. 
For $J_S=0$, $H_K$ realizes the Kitaev spin liquid in the $\tau$ sector, while the $\sigma$ sector remains paramagnetic. For $J_K=0$, $H_S$ realizes a square-lattice spin-orbital liquid. Both solvable limits are characterized by conserved plaquette operators. 
For the present lattice, the elementary plaquettes consist of one hexagon and two inequivalent squares, defined as
\begin{eqnarray}
W_p&=&\mathbb{1} \otimes \tau^x_i\tau^y_j\tau^z_k\tau^x_l\tau^y_m\tau^z_n, \label{eq:WP} \nonumber \\
W_{s}&=&\sigma^z_k\sigma^z_n \otimes \tau^y_i\tau^x_j\tau^y_k\tau^x_n,\label{eq:WS}\\
W_{s'}&=&\sigma^z_k\sigma^z_n \otimes \tau^x_k\tau^y_l\tau^x_m\tau^y_n. \label{WS'} \nonumber 
\end{eqnarray}
Here the site indices in subscript of the plaquette operators run anticlockwise, as shown in Fig.~\ref{fig:lattice_geometry}b. Geometrically, the hexagonal plaquette is tiled by the two square plaquettes, satisfying  $W_p=W_s\, W_{s'}$. Importantly, $W_p$ commutes only with $H_K$ while both $W_s$ and $W_{s'}$ commute with $H_S$. Consequently, the eigenstates of $H_K$ and $H_S$ can be labeled by the eigenvalues of their respective plaquette operators. These plaquette operators square to identity, hence take the eigenvalue $W_{p,s,s'} = \pm 1$.

Before solving the combined Hamiltonian $H_{ks}$, it is convenient to recast them in terms of $\Gamma$-matrices, satisfying the clifford algebra $\{\Gamma^\alpha, \Gamma^\beta\} = 2 \delta^{\alpha \beta}$. We adopt the representation, $\Gamma^\alpha = -\sigma^y \otimes \tau^\alpha$, $\Gamma^4=\sigma^x \otimes \mathbb{1}_2$, $\Gamma^5 = -\sigma^z \otimes \mathbb{1}_2$, along with their antisymmetric products $\Gamma^{\alpha \beta} = i[\Gamma^\alpha, \Gamma^\beta]/2$. Since $\Gamma^1 \Gamma^2 \Gamma^3 \Gamma^4 \Gamma^5 =1$, the orbital pauli matrices $\tau$ can be expressed as $\tau^\alpha =i \Gamma^\alpha \Gamma^4 \Gamma^5 = -\epsilon_{\alpha \beta \gamma} \Gamma^{\beta \gamma}/2$. Then the Eq.~\ref{eqK} and Eq.~\ref{eqSqL} transform to
\begin{eqnarray}
H_{K} &=& \frac{J_{K}}{2}\sum_{\langle ij \rangle_\alpha} \epsilon_{\alpha \beta \gamma}\Gamma_i^{\beta\gamma}\Gamma_j^{\beta\gamma} \label{HKGamma}\\
H_{S} &=& J_{S} \sum_{\langle ij \rangle_{\alpha'}} (\Gamma_i^{\alpha'} \Gamma_j^{\alpha'} + \Gamma_i^{{\alpha'} 5}\Gamma_j^{{\alpha'} 5}).\label{HSGamma} 
\end{eqnarray}
These representations admit an exact Majorana fermion description. For the five Gamma matrices, we have six Majorana fermions at each site, defined as $\Gamma^\zeta = ib^\zeta c$, and $\Gamma^{\zeta \eta} = i b^\zeta b^\eta$, where $\zeta, \eta$ run from 1 to 5. Because the Majorana decomposition enlarges the local Hilbert space, projecting back to the physical subspace requires a local constraint $D = i b^1 b^2 b^3 b^4 b^5 c = 1$. This eliminates the unphysical states via the projection operators  $P = \Pi_j (1 + D_{j})/2$ as $P\ket{\psi}_{\text{total}} = \ket{\psi_{\text{phys}}}$. Relabeling $b^4\rightarrow c^z, \ b^5\rightarrow c^x, \ c\rightarrow c^y $ and utilizing this $D$ identity, Eq.~\ref{HKGamma} and \ref{HSGamma} can be rewritten as
\begin{eqnarray}
H_{K} &=& J_{K} \sum_{\langle ij \rangle_\alpha} iu_{ij}^\alpha c_i^x c_i^y c_i^z  c_j^x c_j^y c_j^z. \label{eq:HKMajorana}\\
H_{S}&=&-J_{S} \sum_{\langle ij \rangle_{\alpha'}} u_{ij}^{\alpha'} (ic_i^x c_j^x +ic_i^y c_j^y).\label{eq:HSMajorana}
\end{eqnarray}
% , the Hamiltonians in terms of Majorana fermions read,}
% \begin{eqnarray}
% H_{K} &=& -J_{K} \sum_{\langle ij \rangle_\alpha} b_i^\beta b_i^\gamma b_j^\beta b_j^\gamma\\
% H_{S}&=&-J_{S} \sum_{\langle ij \rangle_{\alpha'}} u_{ij} (ic_i^x c_j^x +ic_i^y c_j^y).
% \end{eqnarray}
Here $u_{ij}^\alpha = i b_i^\alpha b_j^\alpha$ is defined as the link operators, hence serves as $\mathbb{Z}_2$ gauge field. The plaquette operator defined in Eq. (\ref{eq:WS}) can also be expressed in terms of these link operators as $W=\prod u_{ij}^\alpha$.

% Because the Majorana decomposition enlarges the local Hilbert space, projecting back to the physical subspace requires a local constraint $D = i b^1 b^2 b^3 b^4 b^5 c = 1$. {\color{red}This acts as, $P = \Pi_j (1 + D_{j})/2$ to give back the physical state $P\ket{\psi}_{\text{total}} = \ket{\psi_{\text{phys}}}$.} Utilizing this {\color{red}$D$} identity, the Kitaev term can be rewritten as

% \begin{equation}
%     H_K = J_{K} \sum_{\langle ij \rangle_\alpha} iu_{ij} c_i^x c_i^y c_i^z  c_j^x c_j^y c_j^z. \label{eq:HKMajorana}
% \end{equation}

In the square-lattice limit ($J_K = 0$), the ground state is a spin-orbital liquid residing in the $\pi$-flux sector ($W_{s,s'} = -1$), supporting two flavors of itinerant Majorana fermions $c^x$ and $c^y$, with a low energy Dirac spectrum. Conversely, in the Kitaev limit ($J_S = 0$), the ground state lies in $0$-flux sector ($W_{p} = 1$) with one flavor of itinerant gapless Majorana fermion. A possible gauge choice for this is $u_{\alpha} = 1$, while $u_4 = -1$. Note, while the Majorana Hamiltonian of the square lattice model is bilinear, the Kitaev Hamiltonian involves six fermions. It can be reduced to a bilinear form by defining additional local spin operators, $\sigma^\lambda = c^{\mathsf{T}} L^\lambda c/2$, where $L^\lambda_{\mu \nu} = -i\epsilon^{\lambda \mu \nu}$ are SO(3) generators and $\{\lambda, \mu, \nu\} \in \{x, y, z\}$ in cyclic way. These spin operators also commute with $H_K$, therefore is a conserved quantity with the eigenvalues $\pm 1$. As a result, two Majorana flavours at each site become decoupled, and the remaining Majorana flavour contributes to the quadratic, exactly solvable form of the Hamiltonian. This is why the Lieb's theorem still applies to fix the ground state flux sector for the six-Majorana Kitaev model in Eq. \ref{eq:HKMajorana}.

This exact solvability and the evolution of the phase diagram when both these interactions are simultaneously active remain open questions. It is worth pointing out that, in the intermediate regime when both $J_K$ and  $J_S$ is finite, $\sigma^\alpha$ no longer commutes with $H_S$, destroying the exact solvability. In terms of the plaquette operators in Eq.~\ref{eq:WS},  in this regime $[H_S, W_p] = 0$, but $[H_K,W_{s,s'}]\neq0$ due to the fourth bond. Hence, no static gauge field can be defined on this bond, and $H_S$ in Eq.~\ref{eq:HSMajorana} can be recast as
\begin{align}
H_{S} &= -J_{S} \sum_{\langle ij \rangle_\alpha} u_{ij} (ic_i^x c_j^x +ic_i^y c_j^y) \nonumber\\
&-J_{S} \sum_{\langle ij \rangle_4} (ic_i^z c_j^z) (ic_i^x c_j^x +ic_i^y c_j^y) \label{eq:HSMF}
\end{align}
To solve the combined Hamiltonian  Eq.~\ref{eq:HKS}, a mean-field decoupling into appropriate magnetic and bond channels is required, followed by self-consistent solutions, which we develop in the following sections.

%%%%%%%%%%%%%%%%%%%%
\subsection{Perturbation theory at large $J_K$}\label{subsec:JKPerturbation}
%%%%%%%%%%%%%%%%%%%%

We begin by analyzing the strong Kitaev coupling regime, $|J_K| \gg |J_S|$. As mentioned before, in this limit the ground state is a Kitaev QSL exclusively in $\tau$ sector, while the $\sigma$ sector remains decoupled. Consequently, each spin is free to point in any direction independently, yielding a $2^N$-fold degenerate state for N-site system. The full Kitaev ground-state wavefunction thus takes a direct product form,
\begin{equation}
    \ket{\Psi} =\ket{\psi_K}_\tau \otimes \ket{\prod_i \sigma_i}_\sigma,
\end{equation}
where $|\psi_K\rangle_\tau$ is the ground state of the Kitaev model in $\tau$ sector and $\sigma_i$ is $\ket{\uparrow}$ or $\ket{\downarrow}$ denoting an arbitrary spin configuration on site i. Treating $H_S$ as perturbation, the first order effective Hamiltonian within the ground-state subspace is given by
\begin{align}
H^{\rm eff}
&= P (H_S) P \nonumber \\
&= \sum_{\langle ij \rangle_4} J_S(\sigma_i^x \sigma_j^x + \sigma_i^y \sigma_j^y) \nonumber \\
&\quad + \sum_{\langle ij \rangle_\alpha} J_S
\langle \psi_K | \tau_i^\alpha \tau_j^\alpha | \psi_K \rangle
(\sigma_i^x \sigma_j^x + \sigma_i^y \sigma_j^y) \nonumber \\
&= \sum_{\langle ij \rangle_4} J_S(\sigma_i^x \sigma_j^x + \sigma_i^y \sigma_j^y) \nonumber \\
&\quad + \sum_{\langle ij \rangle_\alpha} J_S
\bigl[-\mathrm{sgn}(J_K)\,0.525\bigr]
(\sigma_i^x \sigma_j^x + \sigma_i^y \sigma_j^y).\label{eq:Heff}
\end{align}
Here $P=|\Psi\rangle \langle \Psi |$ is the projection operator onto the full Kitaev ground state . The value $-\mathrm{sgn}(J_K)\,0.525$ is the exactly calculated nearest neighbour correlation function of the Kitaev Hamiltonian, as illustrated in Ref.~\cite{Baskaran_2007}.  
Equation~\eqref{eq:Heff} shows that an infinitesimal square-lattice spin-orbital coupling lifts the degeneracy of the spin sector and generates an effective XY spin model. 
For $J_K<0$, the effective couplings are unfrustrated, giving ferromagnetic order for $J_S<0$ and antiferromagnetic order for $J_S>0$. 
For $J_K>0$, the fourth bond has the opposite sign relative to the other three bonds, leading to frustrated magnetic interactions.
%Eq.~\ref{eq:Heff} implies that an infinitesimal square term can lead to a magnetic order in the spin sector of the full Hamiltonian, where the nature of the magnetization depends on the sign of $J_K$ and $J_S$. For $J_K < 0$, the ferromagnetic and antiferromagnetic orders emerge for $J_S < 0$ and $J_S > 0$ respectively. However, for $J_K > 0$, the effective coupling on the fourth bond carries the opposite sign to the remaining bonds. As a result, the magnetism becomes frustrated and requires a mean-field calculation.

%%%%%%%%%%%%%%%%%%%%
\subsection{Majorana mean-field theory: $J_K < 0$}
%%%%%%%%%%%%%%%%%%%%

%%%%%%%%%%%%%%%%%%%%
\begin{figure*}[!t]
    \centering
    \includegraphics[width=0.94\textwidth]{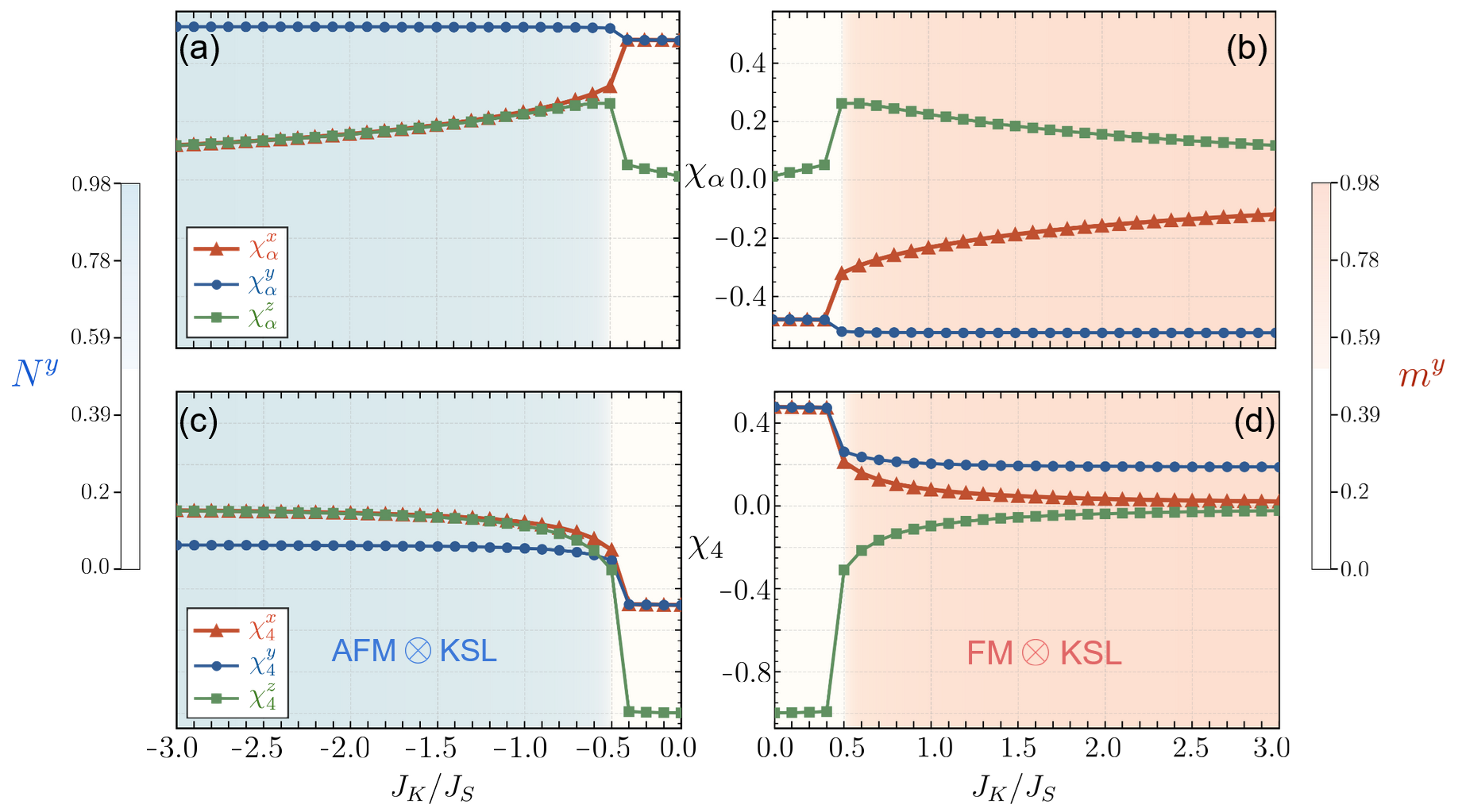}
        \caption{Mean field order parameters across $J_K/J_S$ for $J_K <0$. Results are derived from the self-consistent mean-field solution of Eq.~\ref{eq:HksMF}. Across all panels, the bond order parameters $\chi$ are represented by solid lines, whereas the magnetic order parameters $m$ and $N$ are visualized via the background color maps. (a), (b) shows
        the bond order parameter of 1,2,3 bond $\chi_{\alpha}$ for $J_K/J_S <0$ and $>0$ respectively. Similarly (c), (d) shows the bond order parameter for the fourth bond  $\chi_4$. For $J_K/J_S <0$ in the left column, the system transitions into an antiferromagnetic phase, with the intensity of the blue background indicating the Neel order $N^y$. The right column ( $J_K/J_S >0$) illustrates the emergence of a ferromagnetic phase, where the red background shading denotes the magnetic order parameter, $m^y$.} 
    \label{fig:2}
\end{figure*}
%%%%%%%%%%%%%%%%%%%%

We next study the regime where $J_K$ and $J_S$ are comparable, focusing first on $J_K<0$.  Since the Kitaev term in Eq.~\eqref{eq:HKMajorana} and the fourth-bond part of Eq.~\eqref{eq:HSMF} are interacting (not bilinear) in the Majorana representation, we decouple the Hamiltonian in magnetic and bond channels. The detailed decoupling algebra and the self-consistency equations are presented in Appendix~\ref{Appendix_A}. The decoupling  introduces two types of order parameters; magnetization $m^\lambda_{i}=\langle ic^\mu_{i}c^\nu_{i}\rangle$ and bond order parameter $\chi_{ij}^{\lambda}=\langle ic^\lambda_{i}c^\lambda_{j}\rangle$. $\chi_{ij}^{\lambda}$ measures the fermionic pairing of itinerant $\lambda$ Majorana species across $\langle ij \rangle$ bond. These are analogous to the $\mathbb{Z}_2$ link operators in the exactly solvable limit. The magnetization $m^\lambda_{i}$ captures local polarization in the $\sigma$ sector, which also commutes with $H_K$ in the isolated limit. Physically, a nonzero $m^\lambda_{i}$ corresponds to the development of a spin expectation value $\langle\sigma^\lambda_i\rangle$ on sublattice i, signaling a departure from the paramagnetic phase. Note that, here we have decoupled the magnetic channel only along $x$ and $y$, as the $H_S$ in Eq.~\ref{eq:HSMF} has magnetization terms only in the plane. Following this, the total mean-field Hamiltonian Eq.~\ref{eq:HKS} takes the form

\begin{widetext}
   \begin{align}
       H_{ks}^{MF} = -J_{K}& \sum_{\langle ij \rangle_\alpha} u_{ij}\,  [(m_i^x m_j^x - \chi_{\alpha}^y \chi_{\alpha}^z) (ic_i^x c_j^x)\, + (m_i^y m_j^y - \chi_{\alpha}^x \chi_{\alpha}^z) (ic_i^y c_j^y) - \chi_{\alpha}^x \chi_{\alpha}^y) (ic_i^z c_j^z) \, +  \chi_{\alpha}^y m_j^y (ic_i^z c_i^x) \nonumber\\
        +\, &\chi_{\alpha}^y m_i^y (ic_j^z c_j^x) + \chi_{\alpha}^x m_j^x (ic_i^y c_i^z)\, +   \chi_{\alpha}^x m_i^x (ic_j^y c_j^z)] -J_{S} \sum_{\langle ij \rangle_4} [\chi_4^z(ic^x_ic^x_j +ic_i^y c_j^y) +(ic^z_ic^z_j)(\chi_4^x+\chi_4^y)  \nonumber\\
         -\, & m^y_i(ic^z_jc^x_j)-m^y_j(ic^z_ic^x_i)-m^x_i(ic^y_jc^z_j)-m^x_j(ic^y_ic^z_i)]  -J_{S} \sum_{\langle ij \rangle_\alpha} u_{ij} (ic_i^x c_j^x +ic_i^y c_j^y) + \mathbb{C}_{ks}.\label{eq:HksMF}
   \end{align} 
\end{widetext}
The constant term $\mathbb{C}_{ks}$ is defined as
\begin{align}
    \mathbb{C}_{ks} =&  -J_{K} \sum_{\langle ij \rangle_\alpha} 2 [\chi_{\alpha}^x \chi_{\alpha}^y \chi_{\alpha}^z - \chi_{\alpha}^x m_A^x m_B^x - \chi_{\alpha}^y m_A^y m_B^y] \nonumber\\
 &-J_{S} \sum_{\langle ij \rangle_4}  [m^x_im^x_j+m^y_im^y_j -\chi_4^z(\chi_4^x+\chi_4^y)]. \label{eq:CksMF}
\end{align}

It is important to note that taking  $\mathbb{C}_{ks}$ into account is necessary to correctly identify the true global energy minimum when multiple mean-field solutions coexist at the same parameters.  In the numerical calculations of Eq.~\ref{eq:HksMF} presented here, we set $u_{ij} = 1$ on all three honeycomb bonds, corresponding to the uniform flux-free sector identified as the ground state by Lieb's theorem~\cite{Lieb_1994}.

The self-consistent solution of Eq.~\ref{eq:HksMF} across the entire range of $J_K/J_S$ is plotted in Fig.~\ref{fig:2}. Here, the bond order parameters $\chi$ are plotted with solid lines, while the magnetization $m$ is indicated by the background colors via defining the Neel order parameter $N^y = (m_A^y - m_B^y)/2$ and the magnetic order parameter $m^y = (m_A^y + m_B^y)/2$ for AFM and FM couplings respectively. The magnetic order parameter along $x$ turns out to be zero over the full regime. Evidently, for $|J_K/J_S| \lesssim 0.44$, the system remains in the square spin–orbital liquid phase, with vanishing magnetic order parameters. In this regime, the bond order parameter $\chi_4^z$ saturates to $-1$ (see Fig.~\ref{fig:2}c,d). Notably, $\chi_4^z$ acts as an effective gauge field  to give rise $\pi$-flux sectors as the ground state. Subsequently, we find $\chi_\alpha^z \rightarrow 0$ on the other three honeycomb bonds (Fig.~\ref{fig:2}a,b), confirming that the $z$-Majorana is fully localized and gapped in the spectrum. The equality $\chi^x=\chi^y$ reflects the flavor symmetry between the two itinerant Majorana species of the square-lattice model. This is in conjunction with the isolated square lattice case discussed earlier.  Note that, the signs of $m$ and $\chi$  on all the bonds are determined by the suitable choice of a combination of all mean-field parameters that minimizes the energy of the system.

Near the critical point $|J_K/J_S| \approx 0.44$, at which a first-order phase transition occurs,  $\chi_\alpha^z$ starts acquiring a smaller finite value. Beyond this, the system develops a magnetic order in the $\sigma$ sector. For $J_K/J_S < -0.44$, the system develops an antiferromagnetic order with a finite $N^y$, illustrated by the blue background colour plot in Fig.~\ref{fig:2}(a,c). Similarly, the red colour plot in Fig.~\ref{fig:2}(b,d) illustrates the emergence of ferromagnetic order in the $J_K/J_S > 0.44$ regime, with a finite $m^y$. This is in agreement with our perturbative analysis as explained in sec.~\ref{subsec:JKPerturbation}. It turns out, for both cases in the Kitaev regime (magnetically ordered), the magnetic mean field locally hybridizes the $c^z$ and $c^x$ Majoranas, while the remaining $c^y$ flavor controls the low-energy Dirac spectrum.

To understand the emergence of magnetic order in the spin sector, we recall that the hexagonal plaquette operator $W_p$ in Eq.~\ref{eq:WP} is constructed solely from orbital $\tau$, and the magnetic ordering occurs exclusively in the spin $\sigma$ sector. Therefore, they are conserved throughout the phase diagram, even deep within the magnetically ordered phases. The $\tau$ sector therefore retains its QSL nature, characterized by the $\mathbb{Z}_2$ gauge field and the associated Majorana band structure, even as the $\sigma$ sector orders magnetically. Such a phenomenon, where magnetic and topological order coexist in different degrees of freedom, is known as magnetic fragmentation \cite{Vijayvargia_PRR_2023}. This was first predicted in spin ice systems, where the magnetic moment field can fragment into a fluctuating Coulomb phase coexisting with a magnetic monopole crystal~\cite{Petit2016, SpinIceFrag1, Lefrancois2017, Zorko, Mauws2018, luo2025}. In Kitaev-type spin-orbital models, magnetic fragmentation arises as a transition where local magnetic order coexists with nonlocal topological order \cite{Vijayvargia_PRR_2023}. Typically, the fragmentation in these systems is induced by Heisenberg or Ising interactions that explicitly break the QSL tendency of the parent Hamiltonian. What makes our results interesting is that magnetic fragmentation emerges here entirely from the pure interplay between two distinct spin liquids.

%%%%%%%%%%%%%%%%%%%%
\subsection{Majorana mean-field theory: $J_K > 0$}
%%%%%%%%%%%%%%%%%%%%

Similarly, we now extend our analysis to the $J_K > 0$ regime. As can be seen from the effective Hamiltonian in Eq.~\ref{eq:Heff}, the opposite signs of the couplings on the $\alpha$-bonds relative to the fourth bond introduce frustration, preventing the simple ferromagnetic or antiferromagnetic order. Hence, we perform Luttinger–Tisza (LT) analysis~\cite{LT1, LT2} of the effective spin Hamiltonian in Eq.~\ref{eq:Heff}, to gain insights into the ordering wavevector of the classical ground state and using this to find mean-field solution.

We treat each site as a four-component spinor $(S_i^x, S_i^y, S_j^x, S_j^y)$, where $i \in A$ and $j \in B$ represent the two sublattices and their respective $x$ and $y$ spin components. Fourier transforming the spins as, $S_i^\alpha = N^{-1/2} \sum_k S_k^\alpha e^{i{\bf{k}}.{\bf{r_i}}}$, the momentum space spin Hamiltonian takes the form,
\begin{equation}
\mathcal{H}_{LT}({\bf{k}})  = \left(\begin{matrix}  0& f({\bf{k}}) \cr f^*({\bf{k}})& 0\end{matrix}\right) \otimes \sigma_x
\label{eq:HLT}
\end{equation}

Here $f({\bf{k}}) = J_S [e^{-i{\bf{k}}.{\bf{\delta}}_4}-0.525\, \text{sgn}(J_K) (e^{-i{\bf{k}}.{\bf{\delta}}_1}+e^{-i{\bf{k}}.{\bf{\delta}}_2}+ e^{-i{\bf{k}}.{\bf{\delta}}_3})]$ with the bond vectors defined as, ${\bf{\delta}}_{1,2}=(\pm \sqrt{3}/2, 1/2)$, ${\bf{\delta}}_{3}=(0, -1)$ and ${\bf{\delta}}_{4}=(0, 2)$.

\begin{figure}
    \centering
    \includegraphics[width=1.0\linewidth]{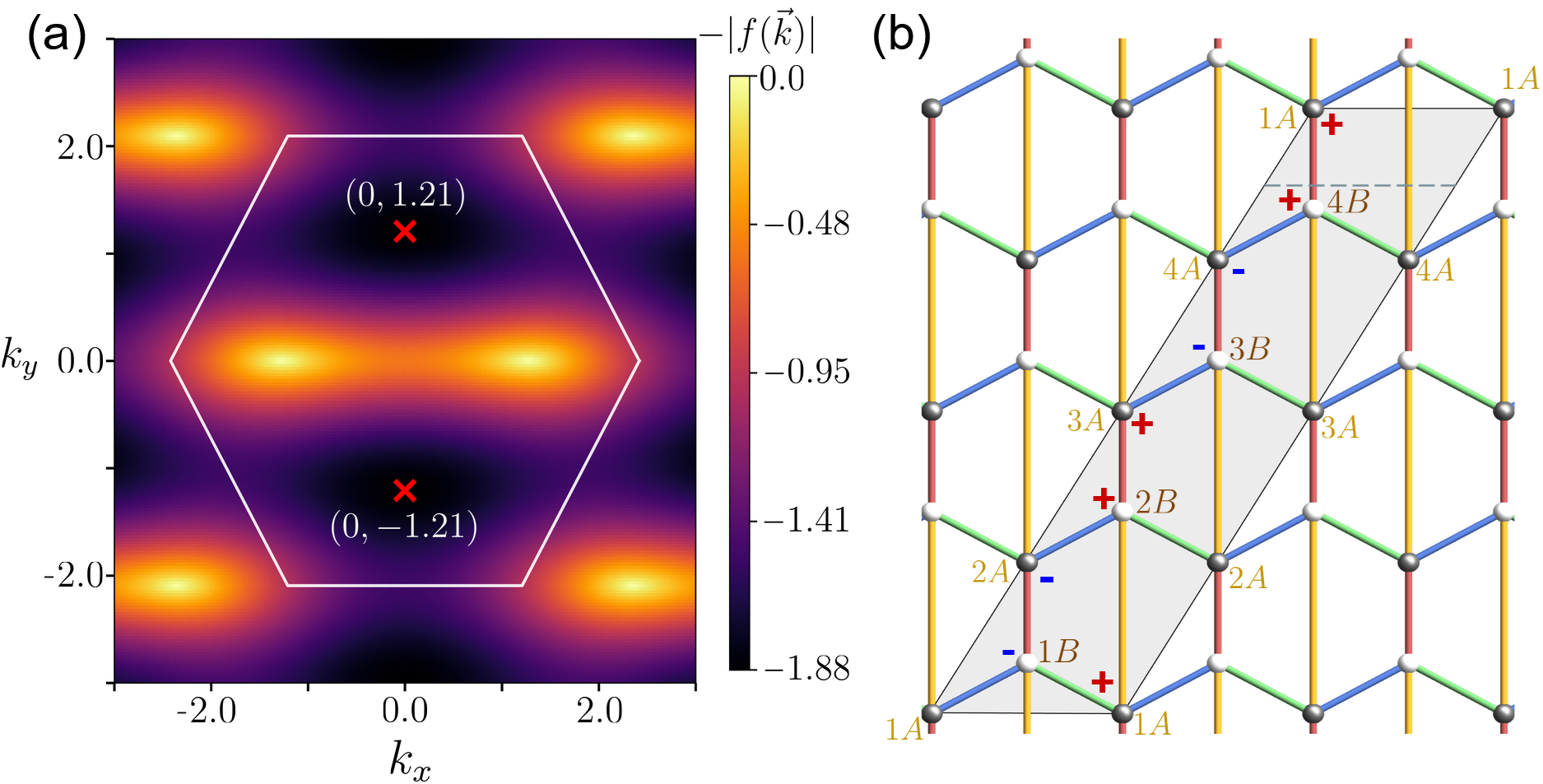}
    \caption{(a) Colourplot of the lowest energy eigenspectrum of $\mathcal{H}_{LT}(\bf{k})$ to find $\bf{k}_{\text{min}}$. The two red crosses indicate the exact coordinates of the optimal ordering wavevectors, ${\bf{k}}_{\text{min}} = (0,\pm 1.207)$. (b) Schematic of the unit cell employed for the self-consistent mean-field analysis in the $J_K>0$. The dashed line represents the exact incommensurate spatial periodicity, $2\pi/{\bf{k}}_{\text{min}}$, determined by LT method. This is approximated by the commensurate 4-site unit cell highlighted in grey. The plus and minus signs on the lattice sites denote the relative signs of the local magnetizations, $m_A^y$ and $m_B^y$, for the spiral antiferromagnetic regime. }
    \label{fig:spiral_lattice}
\end{figure}

As shown in Fig.~\ref{fig:spiral_lattice}(a), the eigenvalue of the Hamiltonian $\pm |f({\bf{k}})|$ yields ordering wavevector of ${{\bf{k}}}_{min} \approx (0, \pm1.21)$.This wavevector is incommensurate with the underlying lattice and therefore indicates a spiral tendency in the classical effective spin model. This is in contrast to the $J_K < 0$ case, where the LT analysis yields ${{\bf{k}}}_{min} = (0, 0)$. As a result, simple ferromagnetic and antiferromagnetic states emerge dictated by the relative phase between the A and B sublattices encoded in the wavefunction.

%%%%%%%%%%%%%%%%%%%%%%%%%%%%%%%%%%%%%%%%
%%%%%%%%%%%%%%%%%%%%%%%%%%%%%%%%%%%%%%%%

As the wavevector is incommensurate, the self-consistent mean-field calculation normally requires an exceptionally large unit cell that can accurately capture the spiral periodicity. To make this manageable, we adopt the minimal commensurate approximation of a four site each cell. This ensures the self-consistency checking to be computationally tractable, also allowing the energy per site to be effectively minimized within a reduced system size, as detailed in Appendix~\ref{Appendix_B}. The unit cell is illustrated in Fig.~\ref{fig:spiral_lattice}(b) shaded in grey.  Accounting the two sublattices and three Majorana flavors per site, this unit cell hosts a total of 64 independent mean-field parameters.

% As the wavevector is incommensurate, the self-consistent mean-field calculation requires a unit cell that can approximately capture the spiral periodicity. We adopt the minimal commensurate approximant of a four site each cell that allows the energy per site to be minimized, as detailed in Appendix~\ref{Appendix_B}. The unit cell is illustrated in Fig.~\ref{fig:spiral_lattice}(b).  Accounting the two sublattices and three Majorana flavors per site, this unit cell hosts a total of 64 independent mean-field parameters. \AV{We can argue this approximation by maybe first stating that a 7 site or 14 site unit cell is better but that would lead to a huge number of MF parameters, therefore we choose 4 site unit cell as that has 64 mean field parameters, which is manageable. }

%%%%%%%%%%%%%%%%%%%%%%%%%%%%%%%%%%%%%%%%
\begin{figure*}[!t]
    \centering
    \includegraphics[width=0.95\textwidth]{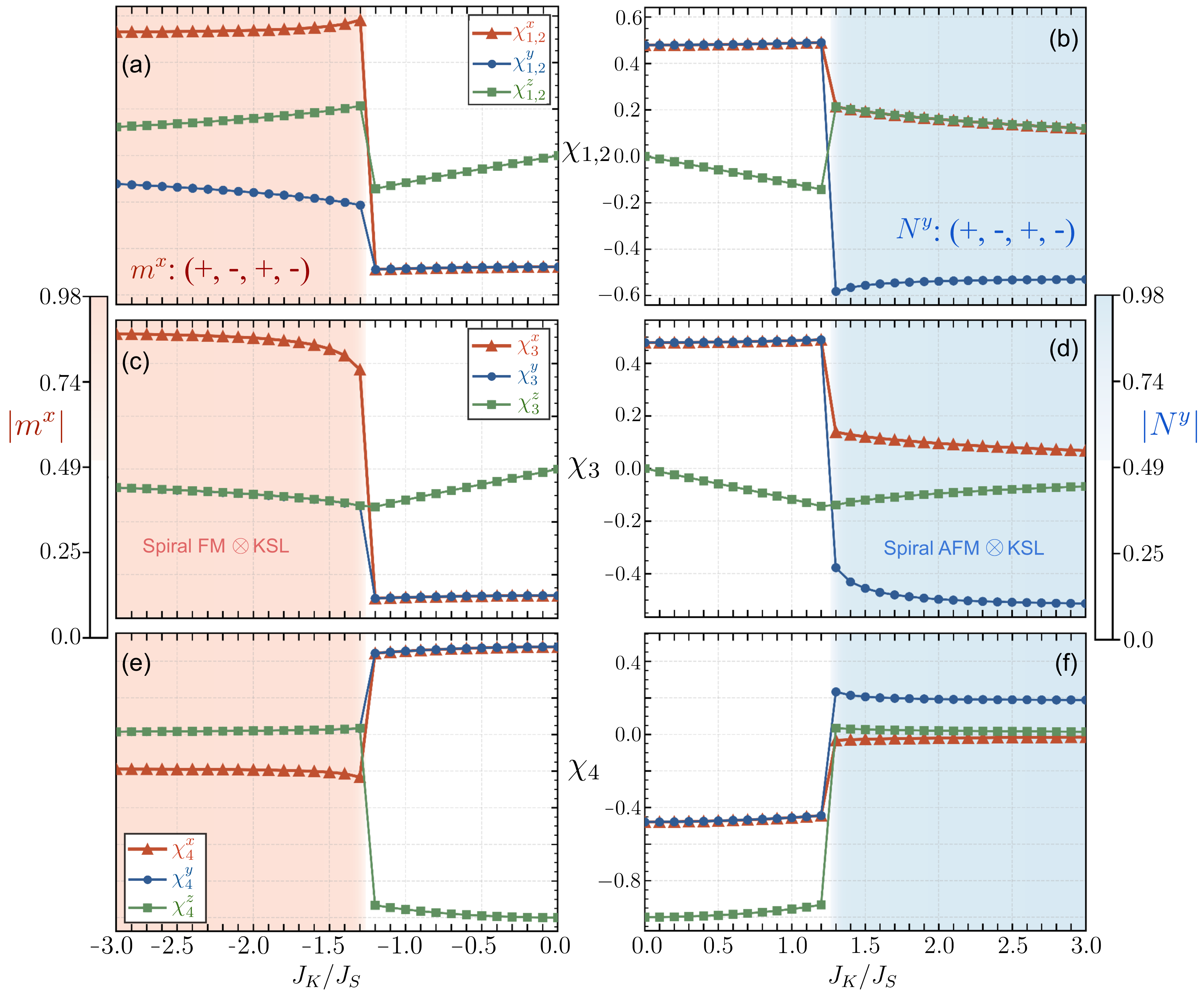}
        \caption{Mean field order parameters across $J_K/J_S$ for $J_K >0$. Results are derived from the self-consistent mean-field solution of Eq.~\ref{eq:HksMF}. Across all panels, the bond order parameters $\chi$ are represented by solid lines, whereas the magnetic order parameters $m$ and $N$ are visualized via the background color maps. (a), (b) shows
        the bond order parameter of 1,2 bond $\chi_{1,2}$ for $J_K/J_S <0$ and $>0$ respectively. Similarly (c), (d) shows the bond order parameter for the third bond $\chi_3$ and (e), (f) for the fourth bond  $\chi_4$. For $J_K/J_S <0$ in the left column, the system transitions into a spiral ferromagnetic phase, with the intensity of the red background indicating the magnitude of the magnetization $|m^x|$. The right column ( $J_K/J_S >0$), illustrates the emergence of a spin antiferromagnetic phase, where the blue background shading denotes the strength of the Neel order parameter, $|N^y|$.}
    \label{fig:3}
\end{figure*}
%%%%%%%%%%%%%%%%%%%%%%%%%%%%%%%%%%%%%%%%
The resulting mean-field phase diagram for $J_K > 0$ is presented in Fig.~\ref{fig:3}. Similar to $J_K < 0$, the square spin-orbital liquid undergoes a first-order phase transition to a magnetically ordered phase. However, here the square spin-orbital liquid remains stable up to higher Kitaev coupling $|J_K/J_S| \approx 1.23$ due to frustration, in contrast to $J_K <0$ regime. For $J_K/J_S \lesssim 1.23$, a spiral ferromagnetic order emerges in the spin sector, illustrated in Fig.~\ref{fig:3}(a,c,e). The magnetic order parameter $m^x$ on the four-site unit cell exhibits an alternating pattern $(+,-,+,-)$. The magnitude of $m^x$ is indicated by the red colorbar on the left of Fig.~\ref{fig:3}. It is worth noting that the superscript $x$ here denotes a chosen direction in the $x$-$y$ plane. As the SO(2) symmetry of square lattice model is already broken in the Kitaev dominated regime for both positive and negative $J_K$, the system can spontaneously prefer any axis, and magnetization vanishes along other direction. As a consequence, the Dirac cones corresponding to the two degenerate Majorana flavors of the square model become gapped, leaving only a single gapless mode aligned with the axis of broken symmetry. Similarly, spiral antiferromagnetic order emerges for $J_K/J_S \gtrsim 1.23$. The magnitude of the Neel order parameter $N^y$ is given by the blue colorbar on the right of Fig.~\ref{fig:3}(b,d,f) with the alternating sign $(+,-,+,-)$ associated with four sites of the unit cell. A schematic of how the sign of $m_A^y$ and $m_B^y$ changes in spiral antiferromagnetic regime is also shown in Fig.~\ref{fig:spiral_lattice}(b). Furthermore, unlike the $J_K<0$ case, here $\chi_3$ behaves differently than $\chi_{1,2}$ due to the alternating intersite local magnetic environment, while 1,2 bonds are on the same site, as observed from Fig.~\ref{fig:spiral_lattice}(b).

% \ccb{While for $J_K <0$ case, finite magnetic order $m^y$ and $N^y$ emerged, here for $J_K >0$, $m^x$ emerges in the spiral ferromagnetic regime. These superscripts on the $m$ denote the specific direction of spontaneous symmetry breaking in the spin sector. As the SO(2) symmetry of the square lattice is broken in this Kitaev regime, the magnetic order parameters can spontaneously choose any specific axis from the $x-y$ plane for both $J_K<0$ and $>0$. Accordingly, the magnetization vanishes in other direction, and the Dirac cones corresponding to the two degenerate Majorana flavors of the square model become gapped, leaving only a single gapless mode aligned with the axis of broken symmetry. Fig.~\ref{fig:3} clearly reflects that, where $\chi^x$ becomes dominant in finite $m^x$ regime and $\chi^y$ in the finite $N^y$ regime. Furthermore, unlike the $J_K<0$ case, here $\chi_3$ behaves differently than $\chi_{1,2}$ due to the alternating intersite local magnetic environment, while 1,2 bonds are on the same site, as observed from Fig.~\ref{fig:spiral_lattice}(b).}

% This is also related to the dominant contribution of a selected Majorana flavour, while the remaining two flavours couple and acquire a mass gap.

In summary, the essential physics that remains in all sectors is the magnetic fragmentation. The spin sector undergoes a symmetry-breaking transition, while the orbital sector retains its $\mathbb{Z}_2$ topological order, manifested in the conserved plaquette flux. Ultimately, the sign of the Kitaev coupling plays a decisive role to induce frustration and determining the ground-state magnetic order, while it will be ferromagnetic, antiferromagnetic, or spiral, as illustrated in Fig.~\ref{fig:lattice_geometry}c. 

% %%%%%%%%%%%%%%%%%%%%%%%%%%%%%%%%%%%%%%%%
% \begin{figure}
%     \centering
%     \includegraphics[width=1.0\linewidth]{Sq_Kitaev_Phase_Diagram.png}
%     \caption{}
%     \label{fig:phase_diagram}
% \end{figure}
% %%%%%%%%%%%%%%%%%%%%%%%%%%%%%%%%%%%%%%%%

%%%%%%%%%%%%%%%%%%%%%%%%%%%%%%%%%%%%%%%%
\section{\label{Section_3} Square - Yao-Lee model}\label{sec:Kitaev_Yao_Lee}
%%%%%%%%%%%%%%%%%%%%%%%%%%%%%%%%%%%%%%%%

%%%%%%%%%%%%%%%%%%%%
\subsection{Model Hamiltonian}
%%%%%%%%%%%%%%%%%%%%

We now replace the orbital Kitaev term by the Yao-Lee interaction while keeping the same square-lattice spin-orbital Hamiltonian and lattice geometry. 
The combined Hamiltonian is $H_{ys}=H_{YL}+H_S$.  The Yao-Lee model is a SO(3)-symmetric generalization of Kitaev's $\mathbb{Z}_2$ spin liquid on honeycomb lattice~\cite{YaoLee2011, Chulliparambil_PRB2021}. In terms of spin  and orbital degrees of freedom, the Hamiltonian is given by
\begin{equation}
    H_{YL} = J_{Y}\sum_{\langle ij \rangle_{\alpha}}(\sigma_i \cdot \sigma_j)(\tau_i^\alpha \tau_j^\alpha). \label{eqYL}\\
\end{equation}

In the $\Gamma$-matrix representation, the Yao-Lee Hamiltonian becomes
\begin{equation}
    H_{YL} = J_{Y} \sum_{\langle ij \rangle_\alpha} (\Gamma_i^\alpha \Gamma_j^\alpha +  \Gamma_i^{\alpha 4}\Gamma_j^{\alpha 4} + \Gamma_i^{\alpha 5}\Gamma_j^{\alpha 5}). 
\end{equation}

Using the same Majorana decomposition as in the previous case, the Yao-Lee Hamiltonian simplifies to
\begin{equation}
    H_{Y} = -J_{Y} \sum_{\langle ij \rangle_\alpha} u_{ij} (ic^x_ic^x_j+ic_i^y c_j^y +ic_i^z c_j^z).
\end{equation}

Thus the Yao-Lee model contains three itinerant Majorana flavors. The plaquette operator $W_p$ defined in Eq.~\eqref{eq:WP} commutes with $H_{YL}$, and the ground state lies in the uniform zero-flux sector. 
In this sector the Majorana bands form triply degenerate Dirac cones. For the combined model, $W_p$ still commutes with $H_{ys}$, while the square plaquettes are no longer conserved. Consequently, the square-lattice term takes the quartic form shown in Eq.~\eqref{eq:HSMF}.

%%%%%%%%%%%%%%%%%%%%
\subsection{Majorana mean-field theory}
%%%%%%%%%%%%%%%%%%%%

To investigate the full phase diagram of $H_{ys}$, we perform mean-field analysis by decoupling the quartic terms of $H_S$ in bond and magnetic channels as detailed in Appendix~\ref{Appendix_A}. Here $H_{YL}$ remains exact due to its quadratic nature. The total mean-field Hamiltonian takes the form
\begin{widetext}
 \begin{align}
   H_{ys}^{MF}& =- \sum_{\langle ij \rangle_\alpha}  [(J_{Y} + J_S)u_{ij}\, (ic_i^x c_j^x + ic_i^y c_j^y) + J_{Y}u_{ij}\,  (ic_i^z c_j^z)] -J_{S} \sum_{\langle ij \rangle_4} [\chi_4^z(ic^x_ic^x_j +ic_i^y c_j^y)+(\chi_4^x+\chi_4^y)(ic^z_ic^z_j)\nonumber\\
   &-\,  m^y_i(ic^z_jc^x_j)-m^y_j(ic^z_ic^x_i)-m^x_i(ic^y_jc^z_j)-m^x_j(ic^y_ic^z_i) + m_i^x m_j^x + m_i^y m_j^y - \chi_4^z (\chi_4^x + \chi_4^y)] .\label{eq:HysMF}
\end{align} 
\end{widetext}

%%%%%%%%%%%%%%%%%%%%%%%%%%%%%%%%%%%%%%%%
\begin{figure}[h]
    \centering
    \includegraphics[width=1.0\linewidth]{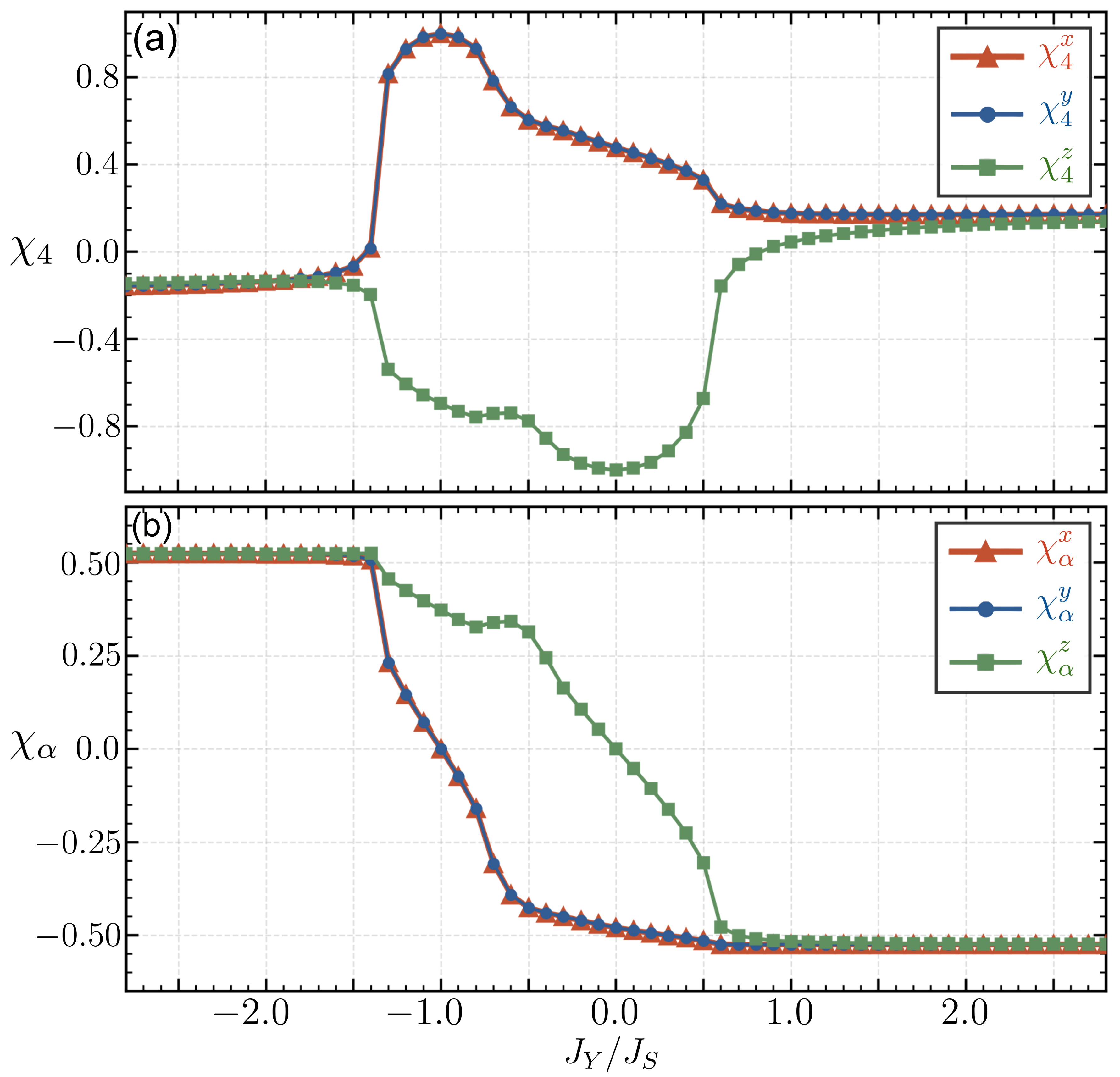}
    \caption{Mean field order parameters as a function of $J_Y/J_S$ obtained by solving Eq.~\ref{eq:HysMF} self-consistently at $J_S = -1$. Figure (a) and (b) depict the behavior of the fourth-bond order parameter $\chi_{4}$ alongside the corresponding bond orders for the 1, 2, and 3 bonds $\chi_{\alpha}$, respectively. Notably, the magnetic order parameters vanish across the entire range of $J_Y/J_S$.}
    \label{fig:5}
\end{figure}
%%%%%%%%%%%%%%%%%%%%%%%%%%%%%%%%%%%%%%%%

\begin{figure*}[!t]
    \centering
    \includegraphics[width=0.92\textwidth]{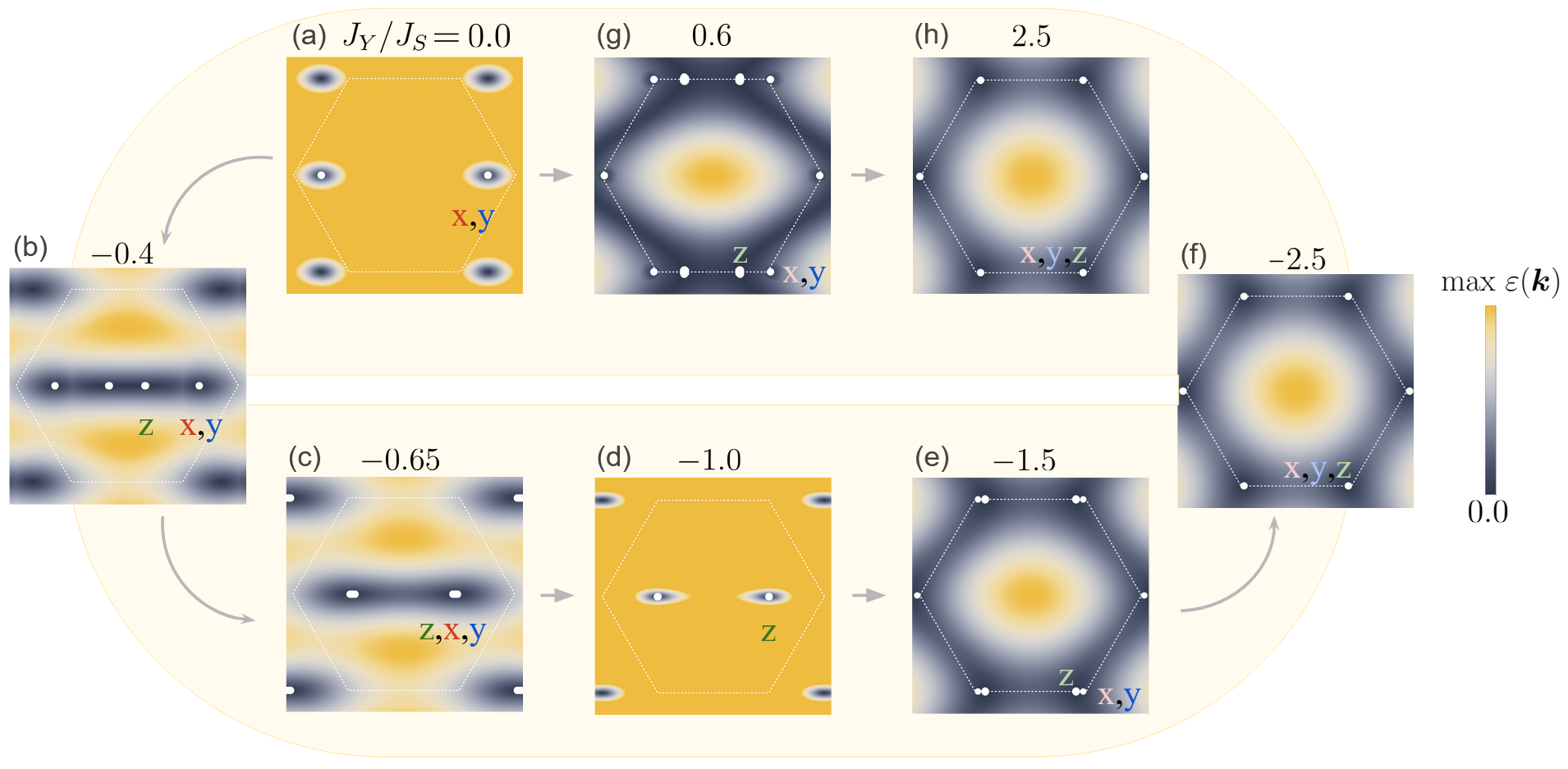}
        \caption{Color plot of the lowest conduction band of the square - Yao-Lee mean-field Hamiltonian (Eq.~\ref{eq:HysMF}) over the full range of $J_Y/J_S$ at $J_S = -1$. The dashed line implies the Brillouin zone boundary, and the white dots depict the position of Dirac points. The figure captures an intricate diagram of Lifshitz transition on how the Dirac points merge and emerge in the Brillouin zone for both (a-f) negative and (g-h) positive regime of  $J_Y/J_S$.}
    \label{fig:6}
\end{figure*}

Eq.~\ref{eq:HysMF} consists of four magnetic and three bond order parameters, which are determined self-consistently. Fig.~\ref{fig:5} illustrates these solutions across the full range of $J_{Y}/J_S$ for $J_S = -1$. The Majorana spectrum in the $J_Y/J_S < 0$ regime reveals some enriching physics. While the square model supports doubly degenerate Dirac cones by $c^x$ and $c^y$ (Fig.~\ref{fig:6}a), at $J_Y/J_S \approx -0.4$, a Lifshitz transition renders another Dirac cone of $c^z$ Majorana near the center of the Brillouin zone (Fig.~\ref{fig:6}b). As observed in Fig.~\ref{fig:6}c, these new Dirac points start moving towards the doubly degenerate Dirac cones and cross them at $J_Y/J_S \approx -0.65$. At $J_Y/J_S \approx -0.75$, the $c^x$ and $c^y$ spectra become gapped, leaving only the $c^z$ Dirac spectrum. Along this trajectory, the point $J_Y/J_S = -1.0$ is special because at this point Fig.~\ref{fig:5} shows $\chi_4^{x,y} = 1$ and $\chi_\alpha^{x,y} = 0$. Therefore $\chi^{x,y}$ act as  static gauge fields on the fourth bond hosting an exactly solvable Hamiltonian. Corresponding Majorana spectrum in Fig.~\ref{fig:6}d shows the position of Dirac cones associated with only the $c^z$ Majorana fermion. The exact solvability and corresponding plaquette operators are discussed in detail later in this section.  Beyond this point, the phase transition accelerates, eventually giving rise to doubly degenerate Dirac spectra near the corner and non-degenerate spectra near the edge (Fig.~\ref{fig:6}e).  At higher $J_Y/J_S$, these Dirac cones merge at the corner (Fig.~\ref{fig:6}f), forming triply degenerate Yao-Lee spin orbital liquid with equal bond order value for all Majorana flavors (Fig.~\ref{fig:5}). 

%%%%%%%%%%%%%%%%%%%%%%%%%%%%%%%%%%%%%%%%
%%%%%%%%%%%%%%%%%%%%%%%%%%%%%%%%%%%%%%%%

For $J_Y/J_S > 0$ region, as $J_Y$ increases, the magnitude of $\chi_4$ decreases, while $|\chi_\alpha|$ increases in other three bonds. Concomitantly, the Dirac cones associated with the $c^x$, $c^y$ Majorana flavors start moving towards the corner of the Brillouin zone. At $J_Y/J_S \approx 0.6$, the Lifshitz transition occurs, and the $c^z$ Majorana flavor forms a new Dirac cone at the Brillouin zone edge. Fig.~\ref{fig:6}g shows two distinct type of Dirac points at $J_Y/J_S = 0.6$, non-degenerate points at the edge, and  doubly degenerate point near the corner. Similar to the $J_Y/J_S < 0$ case, all of them merge at the corner in the Yao-Lee limit (Fig.~\ref{fig:6}h) with equal bond order value as illustrated in Fig.~\ref{fig:5}.

Although exact solvability is lost in the intermediate regime, remarkably the point $J_Y/J_S = -1$  stands as a special exception, where the exact solution is restored. The key lies in the emergence of new conserved quantities as a consequence of opposite coupling strength. Considering $J_S = -1$ and $u_{ij} = 1$, $H_{ys}$ becomes
\begin{equation}
    H_{ys}^{ex} = -\sum_{\langle ij \rangle_\alpha} (ic_i^z c_j^z) + \sum_{\langle ij \rangle_4} (u_4^x + u_4^y)(ic_i^z c_j^z) \label{eq:HysExact}
\end{equation}

 We identify two new mutually commuting link operators $u_4^x = ic_i^x c_j^x$ and  $u_4^y = ic_i^y c_j^y$, which commutes with $H_{ys}^{ex}$ and take values $\pm 1$. This restores the quadratic nature of the Hamiltonian, with $c^z$ Majoranas acting as low-energy excitations. Interestingly, while square lattice model hosts two itinerant Majorana flavours and Yao-Lee model hosts three, due to the exactly opposite coupling strength two of them cancel out. This leaves a single free Majorana flavor in the new solvable Hamiltonian. The ground state of Eq.~\ref{eq:HysExact} corresponds to $u_4^x = u_4^y = 1$, as also observed in Fig.~\ref{fig:5}(a). Consequently, we can define new square plaquette operators,
\begin{eqnarray}
W_{s}^{ex}&=&\sigma^x_k\sigma^x_n \otimes \tau^y_i\tau^x_j\tau^y_k\tau^x_n \label{eq:WSex}\\
W_{s'}^{ex}&=&\sigma^y_k\sigma^y_n \otimes \tau^x_k\tau^y_l\tau^x_m\tau^y_n. \label{eq:WS'ex} \nonumber 
\end{eqnarray}

These operators commute with $H_{ys}^{ex}$. Unlike the standard square lattice model, here $W_{s,s'}^{ex} = 1$ in the ground state due to the negative sign of the fourth bond coupling. While magnetic order emerges in the Kitaev - square model, the mean-field analysis of the Yao-Lee - square model does not give any magnetic order in the $\sigma$ sector. A classical perturbative analysis suggests that for $J_S \gg J_Y$, the magnetic interaction on the decoupled diagonal bonds: $J_S\langle\sigma_i^x \sigma_j^x+\sigma_i^y \sigma_j^y\rangle = 2J_S(S^2-(S^z)^2 - 1)$. Hence it turns out that, for $J_S>0$ the spin singlet state $\ket{0,0}$ is the ground state with $E=-2J_S$ while for $J_S<0$, the $S^z=0$ spin triplet state, $\ket{1,0}$ is the ground state with energy $2J_S$. However, the mean-field analysis in presence of the bond order parameter does not support the emergence of such magnetic ordering.

%%%%%%%%%%%%%%%%%%%%%%%%%%%%%%%%%%%%%%%%
\section{\label{Section_4} Conclusions}\label{sec:Conclusions}
%%%%%%%%%%%%%%%%%%%%%%%%%%%%%%%%%%%%%%%%

In summary, we have investigated the emergent quantum ground state when two distinct exactly solvable spin liquid interactions coexist on a modified honeycomb lattice geometry. For the first model, comprising the Kitaev orbital interaction and square-lattice spin-orbital interaction, perturbative analysis shows that even an infinitesimally small square coupling induces magnetic ordering in the spin sector, while the orbital sector retains its $\mathbb{Z}_2$ quantum spin liquid nature, leading to magnetic fragmentation. In the full parameter regime, we carry out a self-consistent Majorana mean-field analysis. The corresponding solution illustrates that a first-order phase transition drives the system from a pure spin-orbital liquid into ferromagnetic, antiferromagnetic, or incommensurate spiral magnetically ordered states depending on relative strengths of Kitaev and square coupling. For the second model, we combined the Yao-Lee Hamiltonian with the square-lattice Hamiltonian, where the hybrid Hamiltonian remains in the spin-orbital liquid phase. Here, the mean-field solution reveals a rich evolution of the Majorana band structure with multiple Dirac nodes and Lifshitz transitions. Furthermore, when both couplings are equal in magnitude and opposite in sign, the model restores exact solvability with a single itinerant Majorana flavor.

It is worth pointing out that in real materials Heisenberg interactions coexist alongside the bond-dependent interactions considered here. The effects of Heisenberg interactions on related spin liquid models have been studied extensively in the literature~\cite{Urban2020,akram2025, Fornoville2025}, showing an expansion or suppression of the magnetic regime with its relative coupling strength. Furthermore, it is important to move beyond mean-field analysis to capture the true nature of the physics near the phase boundaries. Other promising directions include perturbations that break flux conservation, external field, or even pressure, which may modify orbital composition. Finally, identifying realistic candidate materials and experimentally accessible signatures of magnetic fragmentation would be crucial for connecting these hybrid spin-orbital models to physical realizations.

%%%%%%%%%%%%%%%%%%%%%%%%%%%%%%%%%%%%%%%%
\section{\label{Section_5} Acknowledgement}
%%%%%%%%%%%%%%%%%%%%%%%%%%%%%%%%%%%%%%%%
 ID, KS and AM acknowledge financial support from the Department of Atomic Energy (DAE), Govt. of India, through the project Basic Research in Physical and Multidisciplinary Sciences via RIN4001. ID also acknowledges VIRGO and MPI-PKS, Dresden computing cluster, where most of the numerical calculations were performed. AV and OE acknowledge support from NSF Award No. DMR-2234352.

%%%%%%%%%%%%%%%%%%%%%%%%%%%%%%%%%%%%%%%%
\section{\label{Section_6} Data Availability Statement}
%%%%%%%%%%%%%%%%%%%%%%%%%%%%%%%%%%%%%%%%

The datasets underlying the plots that support this study’s conclusions are available upon reasonable request from the authors.

%%%%%%%%%%%%%%%%%%%%%%%%%%%%%%%%%%%%%%%%
\appendix
%%%%%%%%%%%%%%%%%%%%%%%%%%%%%%%%%%%%%%%%

%%%%%%%%%%%%%%%%%%%%%%%%%%%%%%%%%%%%%%%%
\section{\label{Appendix_A} Self consistent mean field solution}
%%%%%%%%%%%%%%%%%%%%%%%%%%%%%%%%%%%%%%%%

The square lattice model (Eq.~\ref{eq:HSMF}) and the Kitaev model (Eq.~\ref{eq:HKMajorana}) involve four and six Majorana fermion interactions, respectively. To render the quadratic Hamiltonian, we decouple these interactions via mean-field theory in both the magnetic and non-magnetic (bond) channels. We choose the magnetic order parameters, $m^\lambda=\langle ic^\mu c^\nu\rangle $ only along $x$ and $y$, as the square model has magnetic field interaction terms only in the plane. Here, $\lambda, \mu, \nu =\{x,y,z\}$ in a cyclic way. The non-magnetic (Hartree) bond expectation values are defined as $\chi_{\langle ij \rangle_\alpha}^\lambda=\langle ic^\lambda_{i}c^\lambda_{j}\rangle$. Consequently, the square lattice interaction along the fourth bond decouples in the bond channel as
\begin{eqnarray}
    (ic^z_ic^z_j)(ic^x_ic^x_j +ic_i^y c_j^y)=\chi^z(ic^x_ic^x_j +ic_i^y c_j^y)&\\
    +(ic^z_ic^z_j)(\chi^x+\chi^y) -&\chi^z(\chi^x+\chi^y). \nonumber
\end{eqnarray}

Similarly, decoupling the same term in the magnetic channel yields
\begin{eqnarray}
    (ic^z_ic^z_j)(ic^x_ic^x_j +ic_i^y c_j^y)=-m^y_i(ic^z_jc^x_j)-m^y_j(ic^z_ic^x_i)&\\
   -m^x_i(ic^y_jc^z_j) -m^x_j(ic^y_ic^z_i)+m^x_im^x_j+&m^y_im^y_j. \nonumber
\end{eqnarray}
Following the similar procedure, the six-Majorana interactions in Kitaev Hamiltonian are reduced to a quadratic form,

\begin{align}
ic_i^x c_i^y c_i^z c_j^x c_j^y c_j^z &= - \Big[ (m_i^x m_j^x - \chi_{\alpha}^y \chi_{\alpha}^z) (ic_i^x c_j^x)  \\
& + (m_i^y m_j^y - \chi_{\alpha}^x \chi_{\alpha}^z) (ic_i^y c_j^y) - \chi_{\alpha}^x \chi_{\alpha}^y (ic_i^z c_j^z) \nonumber \\
& + \chi_{\alpha}^y m_j^y (ic_i^z c_i^x) + \chi_{\alpha}^y m_i^y (ic_j^z c_j^x) \label{eq:MajoranaDecoupling} \nonumber \\
& + \chi_{\alpha}^x m_j^x (ic_i^y c_i^z) + \chi_{\alpha}^x m_i^x (ic_j^y c_j^z) \nonumber\\
& + 2 (\chi_{\alpha}^x \chi_{\alpha}^y \chi_{\alpha}^z - \chi_{\alpha}^x m_i^x m_j^x - \chi_{\alpha}^y m_i^y m_j^y) \Big] \nonumber
\end{align}

By fixing the gauge to $u_{ij} = +1$ for all $i \in A$ and $j \in B$, the majorana hopping problem possesses translational symmetry on the honeycom lattice. The majorana operators in the Fourier basis takes the form,

\begin{equation}
    c_{j}^\alpha = \sqrt{\frac{2}{N}} \sum_{\mathbf{k} \in BZ/2} ( c_{\mathbf{k}}^\alpha e^{i \mathbf{k}.\mathbf{x}_j} + c_{\mathbf{k}}^{\alpha^{\dag}} e^{-i \mathbf{k}.\mathbf{x}_j})
\end{equation}

While $c_j^{\dag} = c_j$ in real space, in momentum space they obey $c_{\mathbf{k}}^{\dag} = c_{-\mathbf{k}}$. The full mean-field Hamiltonian for $J_K < 0$ regime can then be constructed using the six-component spinor basis $(c_A^x, c_A^y, c_A^z, c_B^x, c_B^y, c_B^z)$ as,
\begin{equation}
\mathcal{H}^{\mathrm{MF}}_{ks}(\mathbf{k})=
\begin{pmatrix}
\mathcal{H}_A & \mathcal{H}_{AB} \\
\mathcal{H}_{BA} & \mathcal{H}_B
\end{pmatrix}.
\end{equation}

The intra-sublattice components are expressed in terms of  $L^\lambda_{\mu \nu} = -i\epsilon^{\lambda \mu \nu}$ as,
\begin{equation}
\begin{aligned}
\mathcal{H}_{A} &=
\left( 2 J_K m_B^x \sum_{\alpha} \chi_{\alpha}^x - 2 J_S m_B^x \right) L^x \\
&\quad +
\left( 2 J_K m_B^y \sum_{\alpha} \chi_{\alpha}^y - 2 J_S m_B^y \right) L^y .
\end{aligned}
\end{equation}

The block $\mathcal{H}_{B}$ is obtained simply by interchanging the A and B sublattice indices. The inter-sublattice block takes the diagonal form,
\begin{equation}
\mathcal{H}_{AB}(\mathbf{k})=
\begin{pmatrix}
D_x(\mathbf{k}) & 0 & 0 \\
0 & D_y(\mathbf{k}) & 0 \\
0 & 0 & D_z(\mathbf{k})
\end{pmatrix}.
\end{equation}

Here,
\begin{eqnarray}
D_x(\mathbf{k})&=& G_x - 2iJ_S\Big[\sum_\alpha f_\alpha \bf(k) + \chi_4^z g \bf(k)\Big]  \nonumber \\
D_y(\mathbf{k})&=&G_y - 2iJ_S\Big[\sum_\alpha f_\alpha \bf(k) + \chi_4^z g \bf(k)\Big] \nonumber\\
D_z(\mathbf{k})&=&G_z -2iJ_S (\chi_4^x+\chi_4^y)g\bf(k) \\
  \nonumber
\end{eqnarray}

for $G_{\lambda} =  -2iJ_K\sum_\alpha(m_A^\lambda m_B^{\lambda}- \chi_\alpha^\mu \chi_\alpha^\nu) f_\alpha\bf(k)$. Here, the bond dependent phase factors take the form, $f_\alpha\bf(k) =e^{i{\bf{k}}.{\bf{\delta}}_\alpha}$ and $g\bf(k) =e^{i{\bf{k}}.{\bf{\delta}}_4}$. In total, this mean-field Hamiltonian is governed by 4 magnetic order parameters and 12 bond order parameters. We solved this Hamiltonian self-consistently via an iterative procedure on a $48 \times 48$ momentum-space grid, achieving rigorous convergence up to a tolerance of $10^{-15}$.

It is important to mention that, for $J_K>0$ regime, as the unit cell increases, corresponding $\bf{k}$-space Hamiltonian will be make of 24 component spinor, for which we performed a similar self-consistent solution over 64 mean field parameters.

%%%%%%%%%%%%%%%%%%%%%%%%%%%%%%%%%%%%%%%%
\section{\label{Appendix_B} Unit cell consideration for $J_K>0$}
%%%%%%%%%%%%%%%%%%%%%%%%%%%%%%%%%%%%%%%%

%%%%%%%%%%%%%%%%%%%%%%%%%%%%%%%%%%%%%%%%
\begin{figure}[!t]
    \centering
    \includegraphics[width=1.0\linewidth]{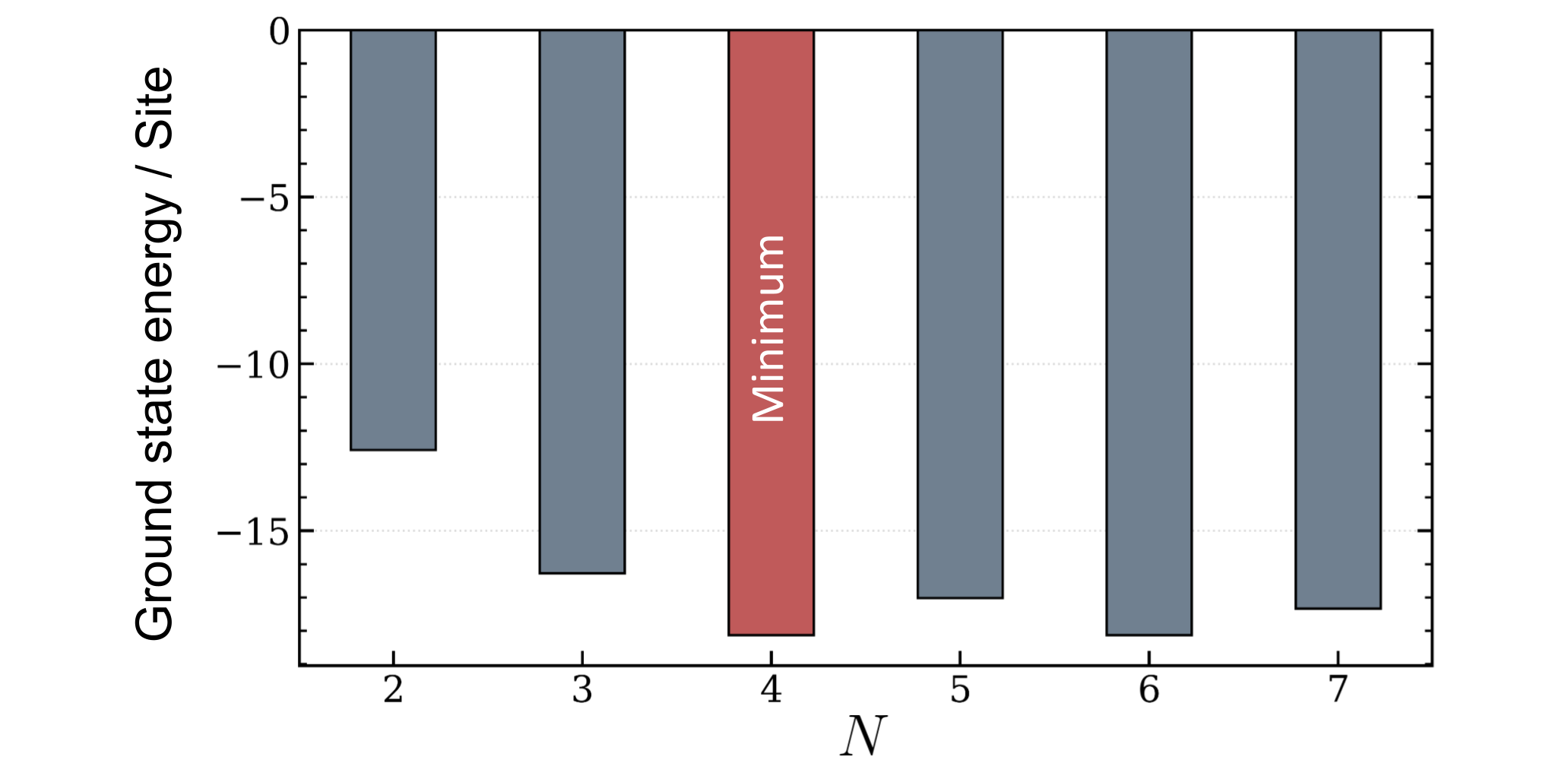}
    \caption{The minimum energy per site for a self-consistent mean field solution at $J_K = 2$ and $J_S = 1$. The figure shows $1 \times 4$ site unit cell gives the minimum energy per site, which we considered for all calculations at $J_K > 0$.}
    \label{fig:7}
\end{figure}
%%%%%%%%%%%%%%%%%%%%%%%%%%%%%%%%%%%%%%%%

Since the ordering wavevector for the Kitaev - Yao-Lee model in $J_K > 0 $ is incommensurate, a commensurate unit cell approximant must be employed for the mean-field treatment. The classically determined wavevector, ${{\bf{k}}}_{min} \approx (0, \pm1.21)$ implies that the magnetic structure is uniform along the $x$-axis but modulated along the $y$-axis, with the modulation period lying between three and four lattice constants. Therefore, in order to find the best supercell, we systematically examined $1\times N$ unit cells for $N =1$ to $7$ and optimized $16N$ mean-field parameters for each case. Note that, when $N$ is large, there exists an extensive number of solutions corresponding to local minima. To mitigate getting trapped in local minima, we started the self-consistency loop from $1000$ different randomly chosen starting points. Fig.~\ref{fig:7} shows that the ground-state energy per site is minimized for $N=4$. Thus, we chose a four-site unit cell as the best commensurate approximant within the manageable regime for our calculations across the entire parameter regime.

% \textcolor{red}{AM: Do we choose the case with minimum energy among them? By itself such an approach is not a guarantee of global minima. We can rephrase as "We choose the following strategy to mitigate getting trapped in local minima. We started with...as written above with a comment that the lowest energy case is chosen"}

\bibliography{references}

% ---------- End ----------

\end{document}